\documentstyle[12pt,epsf]{article}

\newcommand{\al}{\alpha}
\newcommand{\partderiv}[2]{\frac{\partial #1}{\partial #2}}
\newcommand{\sd}{\mathrel{\subset\hspace{-2.3ex}\times}}
 
\newcommand{\M}{{\cal M}}
\newcommand{\N}{{\cal N}}
\newcommand{\X}{{\cal X}}
\newcommand{\Hcal}{{\cal H}}
\newcommand{\Ocal}{{\cal O}}
 
\begin{document}
 
\title{Path Group in gauge theory and gravity\thanks{Reported at the 
XXIV International Colloquium on Group Theoretical Methods in Physics 
(ICGTMP-2002 or Group 24), Paris, July 15-20, 2002.}}
\author{Michael B. Mensky\\ 
{\normalsize P.N. Lebedev Physics Institute, Moscow,Russia}}
\date{}

\maketitle

\begin{abstract}
Applications of the Path Group (consisting of classes of continuous 
curves in Minkowski space-time) to gauge theory and gravity are 
reviewed. Covariant derivatives are interpreted as generators of an 
induced representation of Path Group. Non-Abelian generalization of 
Stokes theorem is naturally formulated and proved in terms of paths. 
Quantum analogue of Equivalence Principle is formulated in terms of 
Path Group and Feynman path integrals. 
\end{abstract}

\newpage

\tableofcontents

\section{Introduction}

One of the most important concepts in the modern quantum field theory 
is gauge field. This concept was introduced \cite{YM} on the basis of 
gauge symmetry, i.e. invariance under localized (depending on 
space-time points) symmetry groups. Later it became clear that gauge 
theory may be naturally formulated on the basis of such mathematical 
formalism as connections in fiber bundles and their curvatures. 
Gravity was from the very beginning formulated as theory of curved 
(pseudo-)Riemannian spaces. Fiber bundles turned out to be also 
efficient mathematical formalism for gravitational fields. 

Here we shall give a short survey of an alternative mathematical 
background for both gauge theory and gravity, namely Path Group 
(\cite{MBM72,MBM74,MBM78,MBM79}, see \cite{MBM83,Warsaw} for reviews). 
Path Group (PG) is a generalization of translation group differing 
from the latter in that it may be applied to particles in external 
gauge and gravitational fields. 

Formally Path Group may be defined as a set of certain classes of 
continuous curves in Minkowski space (generalization on the case of 
paths in an arbitrary group space is possible). Concept of PG came up 
as development of the Suvegesh's groupoid of parallel transports 
\cite{Suvegesh,Suvegesh2} (a groupoid differs from a group in that not any pair 
of its elements may be multiplied). The goal was to find a universal 
group such that parallel transports in various space-times be its 
representations. 

The concept of PG may be obtained also in the attempt to globalize 
infinitesimal translations of a tangent space to a curved space-time. 
PG arises then instead of the usual translation group since curvature 
makes translations in different directions not commutative (see 
\cite{Jackiw,Jackiw2,Jackiw3,Jackiw4} for other types of non-commutative translations).

PG reduces geometry to algebra: various geometries (gauge and 
gravitational fields) are nothing else than representations of the 
universal PG. In case of gauge fields the representation is 
simpler in that Lorentz group may be factorized out. In case of 
gravitational fields both Lorentz and PG (united to give the 
generalized Poincar\'{e} group) essencially participate in the 
constructions. 

The natural character of PG is seen from the facts that it allows 
1)~to give a group-theoretical interpretation of covariant derivatives 
(both for gravitational or/and gauge fields) as generators of relevant 
representations of PG, 2)~to formulate and prove a non-Abelian version 
of Stokes theorem and 3)~to reduce the path integral in a curved 
space-time to the path integral in the flat space (a quantum version 
of Equivalence Principle)

\section{Path Group}\label{Sect-PG}

An element of Path Group (PG) is defined as a {\it class of curves} in 
Minkowski space constructed in such a way that the classes form a 
group. For curves in Minkowski space, $\{ \xi \} = \{ 
\xi(\tau)\in{\cal M}\}$, operation of multiplication $\{ \xi' \}\{ \xi 
\}$ may be naturally defined as passing of two curves one after 
another and inversion $\{ \xi \}^{-1}$ as passing the same curve in 
the opposite direction. However, the set of all continuous curves is 
not, in respect to these operation, a group: 1)~not all pairs of 
curves may be multiplied to give a continuous curve, 2)~multiplication 
(when defined) is not assotiative and 3)~product of a curve by 
its inverse does not yield a unit element (i.e. such one that 
multiplication by it does not change an arbitrary curve). 

To correct the second defect, one may consider differently 
parametrized curves to be equivalent (and include them in the same 
class). To correct the third defect, one may consider equivalent those 
curves which differ by inclusion of `appendices' of the form 
$\{\xi\}^{-1}\{\xi\}$ (and go over to else wider classes of curves). 
At last, the first defect is overcome if we include in the same class 
those curves which differ by general shift: $\xi'(\tau)=\xi(\tau)+a$ (see 
Figure).    
\begin{figure}[ht]
\let\picnaturalsize=N
\def\picsize{1.7in}
\ifx\nopictures Y\else{\ifx\epsfloaded Y\else\input epsf \fi
\let\epsfloaded=Y
{\hspace*{\fill}
 \parbox{2in}{\ifx\picnaturalsize N\epsfxsize \picsize\fi  
\epsfbox{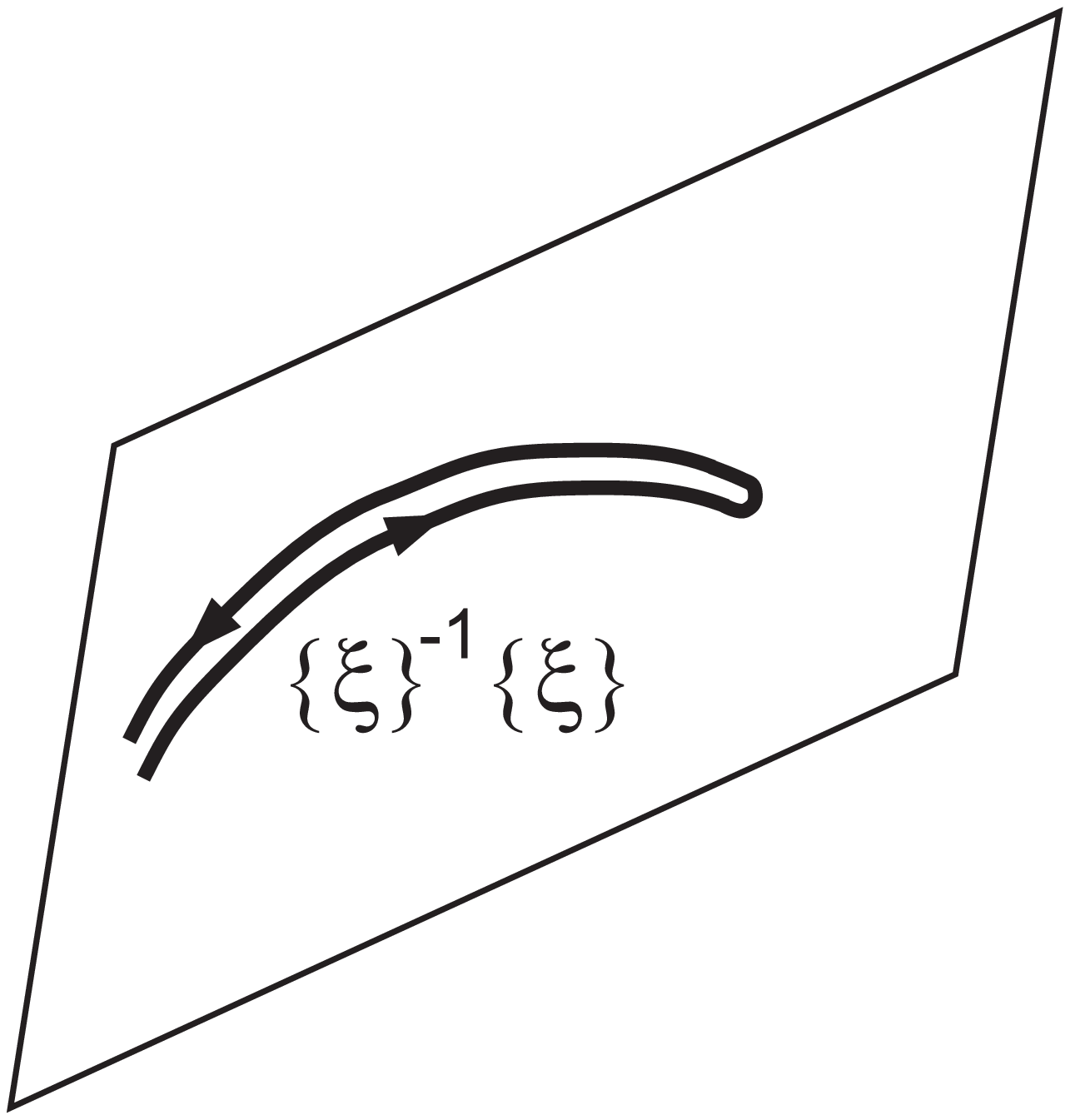}}\hfill
 \parbox{2in}{\ifx\picnaturalsize N\epsfxsize \picsize\fi  
\epsfbox{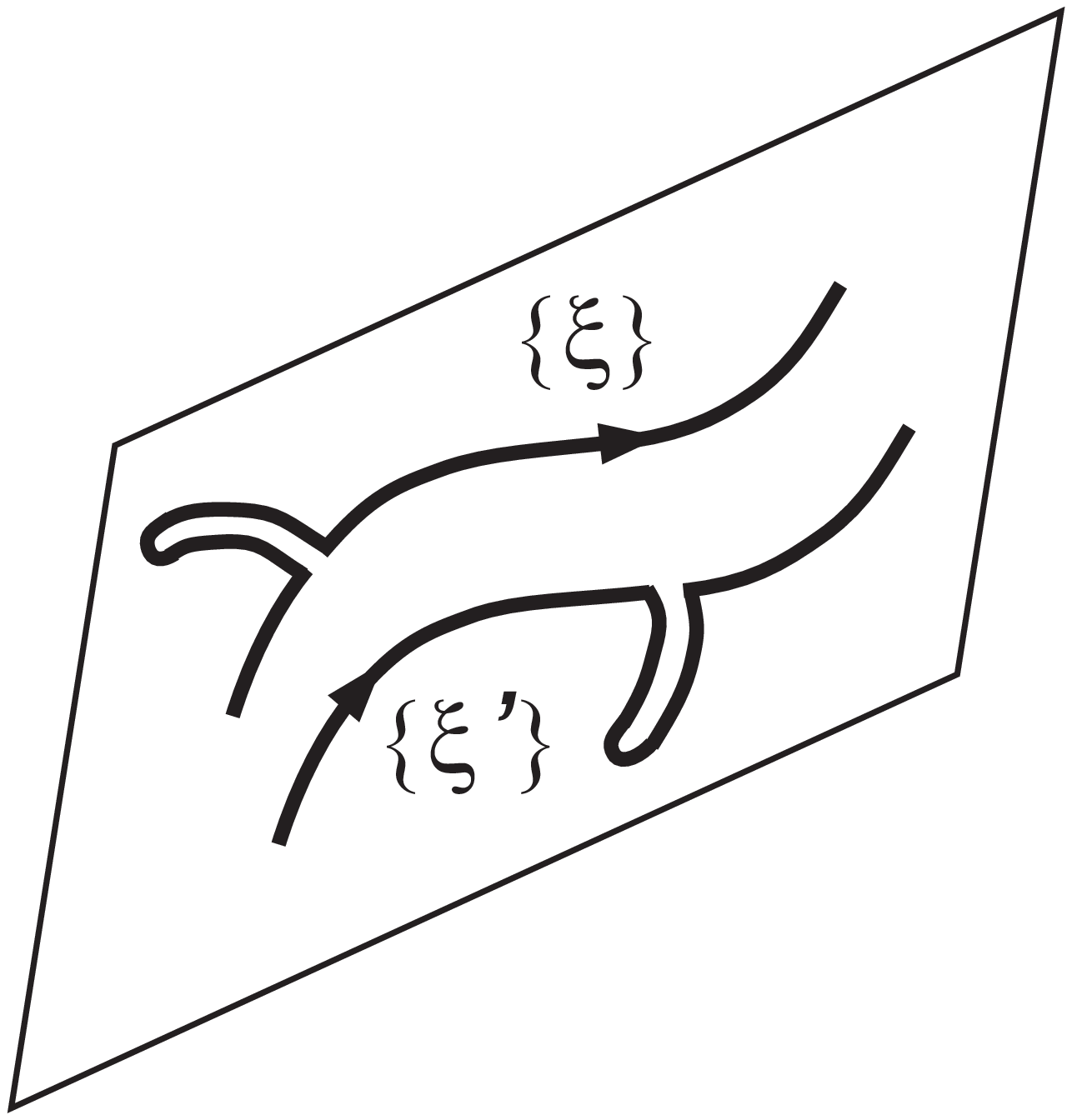}}\hspace*{\fill}}}\fi\\
\hspace*{\fill}$\mbox{`Null' curve}$\hfill \quad 
\hfill $\mbox{Equivalent curves $\{\xi'\}=\{\xi\}$}$\hspace*{\fill}\\
\label{Fig-gr-paths}
\end{figure}

The resulting class of curves is called a {\it path} and denoted by 
$p$ or $[\xi]$. All paths form {\it Path Group} $P$. A path may be 
presented by any curve from the corresponding class. 

If the curves which differ by general shift are not considered 
equivalent, we have more narrow classes. The end points are the same 
for all curves in such a class. It may be therefore called a {\it 
pinned path} and denoted as $\hat{p}=p_x^{x'}=[\xi]_x^{x'}$ (or $p_x$) 
where $x$ is the initial and $x'$ final point of any curve presenting 
the given pinned path. 

All pinned paths form a groupoid $\hat{P}$ since not any pair of 
pinned paths may be multiplied. Pinned paths are often convenient for 
constructing representations of the group $P$ of free paths. It is 
important that the pair $p,x$ of a free path and a point unambiguously 
determines the pinned path $p_x$ starting in $x$ and having the same 
shape as $p$. 

Paths may be defined \cite{MBM83,Warsaw} for an arbitrary group space 
$G$ leading to the group of paths $P(G)$ in $G$ (the role of general 
shift is played in this case by right shift in the group). We shall 
restrict ourselves by considering only the path group in Minkowski 
space, $P=P(\M)$. This will prove to be sufficient for 
applications to gauge theory and gravity. 

\section{Gauge fields as representations of PG}

Theory of free elementary particles is governed by 
the group of translations. Starting from Path Group 
instead of the translation group we obtain theory of particles in 
an external gauge field. Gauge fields thus arise independently of the 
idea of gauge symmetry. 

The key point for constructing theory of particles is that Path Group 
$P$ (generalized translations) acts on Minkowski space $\M$ 
transitively. An important consequence is that particles have to be 
described by the induced representations of Path Group. 

\subsection{Group and localization}

Theory of elementary particles may be constructed starting from a 
certain group and implying the requirement of locality 
\cite{GroupQuant,GroupQuant2,MBM83} (see \cite{Rowe} for an analogous 
construction). In our case Path Group $P$ will play the role of the 
governing group and Minkowski space $\M$ the role of the localization 
space. Note that the action of the group $P$ on the space $\M$ is 
naturally defined as shifting a point $x\in\M$ along the path $p\in 
P$ (see Figure).  
\begin{figure}[ht]
\let\picnaturalsize=N
\def\picsize{2in}
\ifx\nopictures Y\else{\ifx\epsfloaded Y\else\input epsf \fi
\let\epsfloaded=Y
{\hspace*{\fill}
 \parbox{2in}{\ifx\picnaturalsize N\epsfxsize \picsize\fi  
\epsfbox{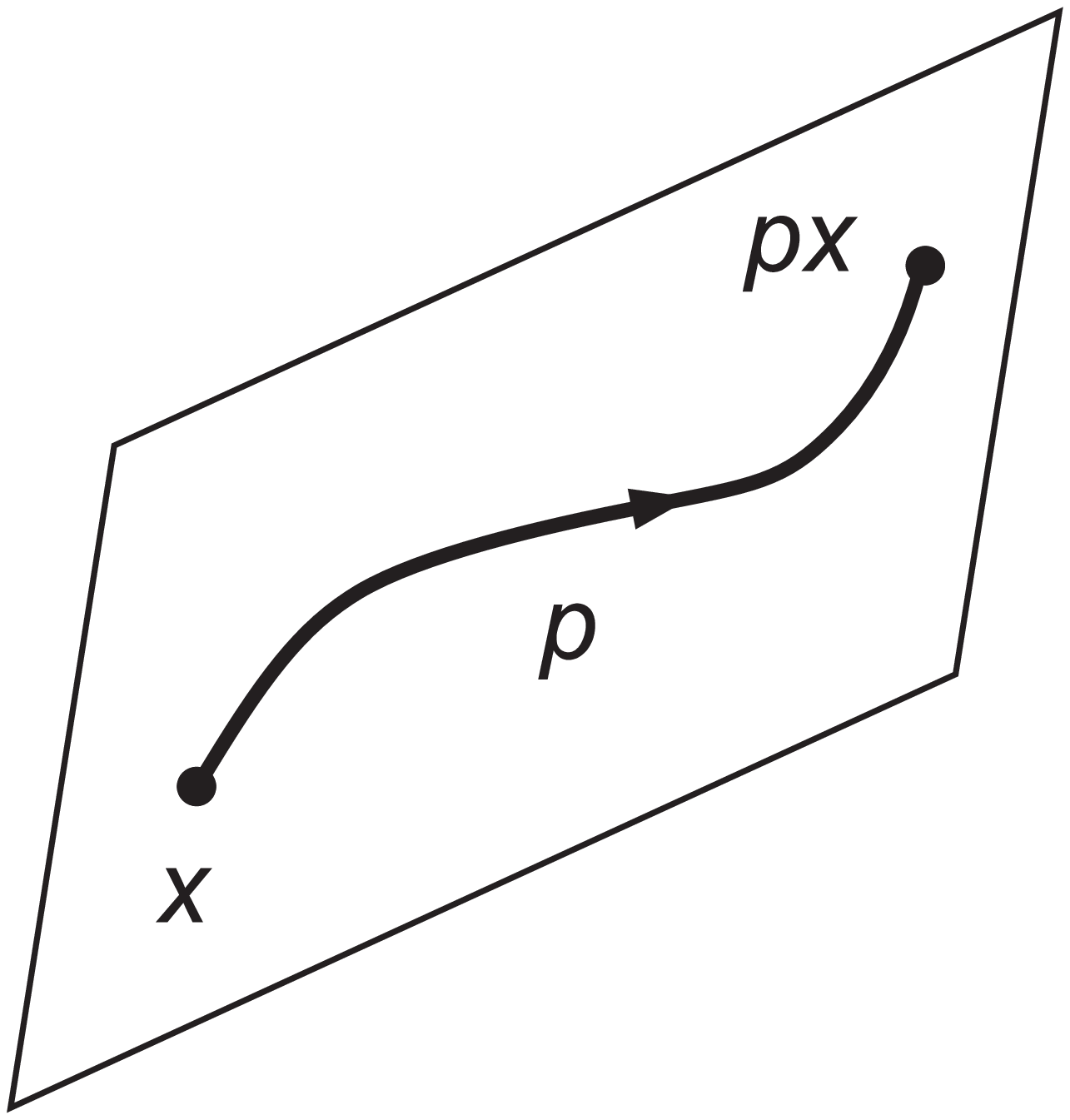}}\hfill
 \parbox{2in}{\ifx\picnaturalsize N\epsfxsize \picsize\fi  
\epsfbox{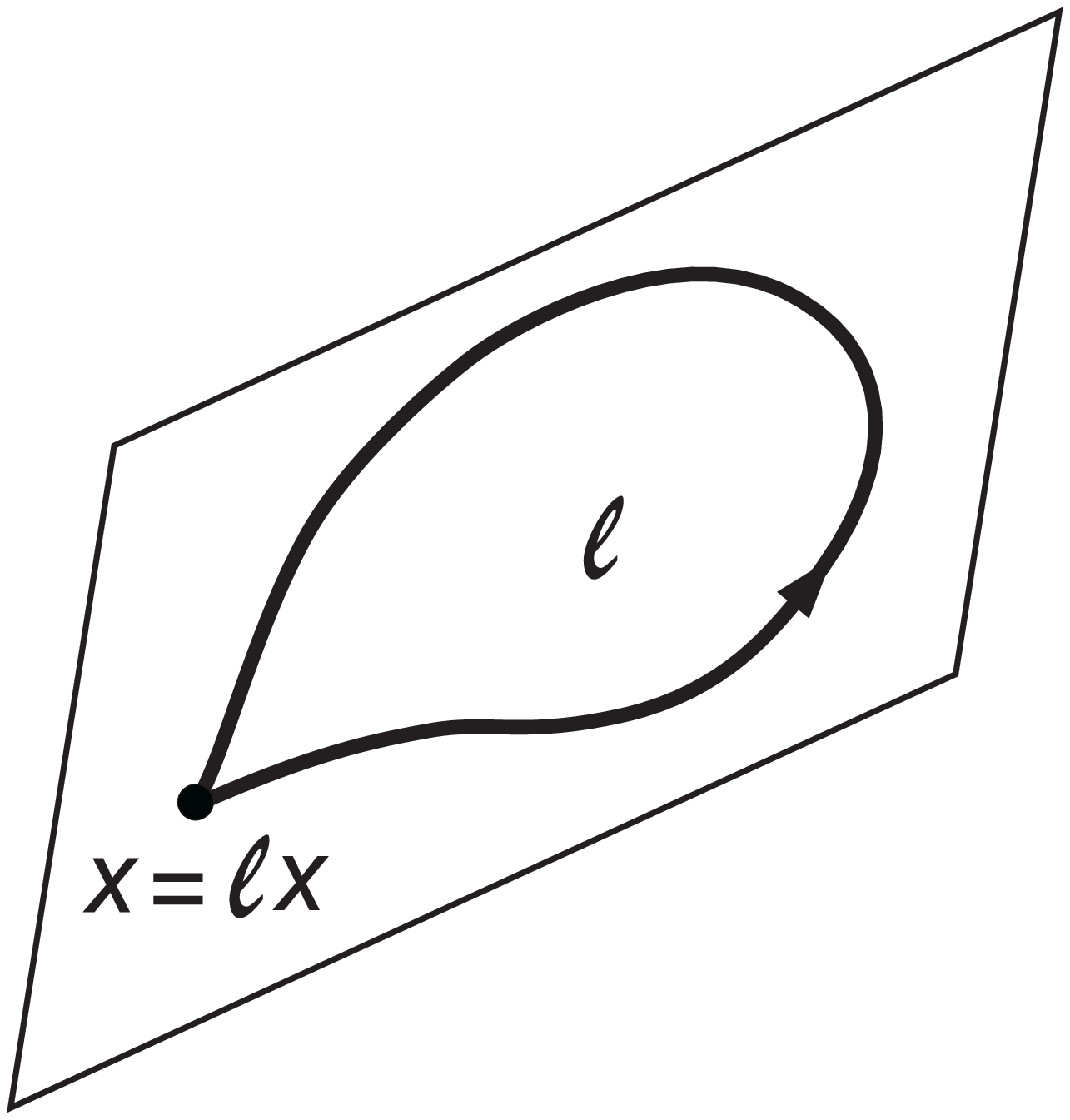}}\hspace*{\fill}}}\fi\\
\hspace*{\fill}$\mbox{Path acting on a point}$\hfill \quad 
\hfill $\mbox{Invariance of points under action of loops}$\hspace*{\fill}\\
\label{Fig-act-Mink}
\end{figure}
An arbitrary point in $\M$ is invariant under the action of 
closed paths (loops). Therefore, the subgroup of loops 
$L\subset P$ is a stabilizer of an arbitrary point. This means that 
the space $\M$ may be presented as a quotient space $\M=P/L$. 

If some point $\Ocal\in\M$ is chosen as an origin in $\M$, then the 
point $x\in\M$ is identified with the coset $p'L\in P/L$ where 
$x=p'\Ocal$. The space $\M=P/L$ will serve in our case as a 
localization space. 

The general scheme of constructing theory of particles starting 
from the group and localization space is following 
\cite{GroupQuant,GroupQuant2,MBM83}. 

Let theory of particles be governed by the group $P$ (which may be 
symmetry or kinematic group) and characterized by localization in the 
space $\M$ (i.e. the subspace of states $\Hcal_x$ is defined in which 
the particle is localized in an arbitrary point $x$ of $\M$). Then the 
group $P$ must act on the space $\M$ (so that $p:\; x\rightarrow px$ 
for $x\in\M$, $p\in P$), and this action must be in accord with the 
representation $U$ of $P$ acting in the space of the particle's 
states: 
$$
U(p)\Hcal_x=\Hcal_{px}. 
$$
A representation $U(P)$ possessing this property is called 
imprimitive. 

Assume that the action of $P$ on $\M$ is transitive (this may 
be done without loss of generality, because otherwise we may divide 
$\M$ in imprimitive subspaces). Then $\M$ is a homogeneous space and 
can be presented as a quotient space, $\M=P/L$, with an appropriate 
subgroup $L\subset P$. 

The representation $U(P)$ acts then transitively on the set of 
subspaces $\{ \Hcal_x|x\in\M=P/L\}$. According to the imprimitivity 
theorem \cite{Coleman}, such a representation (transitive but 
imprimitive) is equivalent to the representation induced from some 
representation $\al(L)$ of the subgroup $L$: 
$$
U(P)=\al(L)\uparrow P.
$$

This gives a receipt for constructing theory of local particles. 
Knowing the group $P$ and the space $\M=P/L$ of localization we can 
restore the representation $U(P)=\al(L)\uparrow P$ acting in the space 
of states $\Hcal$ of the particle (and therefore can restore the space 
$\Hcal$ itself). To do this, we have 1)~to choose arbitrarily a 
representation $\al(L)$ of the subgroup $L$ and 2)~to induce it onto 
the whole group $P$. 

Inducing a representation $\al(L)$ of a subgroup onto the whole group 
$P$ may be achieved in the following way \cite{Coleman}. Vectors of 
the carrier space $\Hcal$ of the induced representation 
$U(P)=\al(L)\uparrow P$ are presented by functions $\Psi(p)$ on $P$ 
with values in the carrier space ${\cal L}_{\al}$ of $\al(L)$ with the 
additional {\it structure condition} imposed on the functions: $$ 
\Psi(pl)=\al(l^{-1})\Psi(p), \quad l\in L, p\in P $$ The induced 
representation acts on these functions by left shifts: $$ 
(U(p)\Psi)(p')=\Psi(p^{-1}p'), \quad p,p'\in P. $$

The representation $\al(L)$ is arbitrary in this construction. It 
describes {\it internal degrees of freedom} of the particle which 
it possesses even if its localization in $\M$ is fixed. 
States of the particle are described by the functions $\Psi(p)$ 
depending on elements of the group $P$. In our case these are paths, 
so we arrive to the non-local formalism of path-dependent wave 
functions. It is a group-theoretical version of Mandelstam's 
path-dependent fields \cite{Mandelstam,Mandelstam2,Bialynicki}. 

However, despite of the non-local form, the theory must be essentially 
local because the requirement of locality was imposed from the very 
beginning. Therefore, {\it explicitly local form} of the 
representation must exist.  

For constructing this form we need an extension $\al(P)$ of the 
representation $\al(L)$ onto the group $P$ such that 
$\al(pl)=\al(p)\al(l)$ for arbitrary $p\in P$ and $l\in L$. Local wave 
functions $\psi(x)$ may be defined then as 
$$
\psi(x)=\al(p')\Psi(p') 
$$
where the point $x\in\M$ corresponds to the coset $p'L\in P/L$.
Now the action of the induced representation is given by 
$$
(U(p)\psi)(x)=\al(p')[\al(p^{-1}p')]^{-1}\psi(p^{-1}x)
$$
(for simplicity we denote this representation by the same letter 
although it is only equivalent but not identical to the preceding 
one). 

\subsection{Particles in gauge field}

Applying the general scheme to the Path Group $P$, its subgroup of 
loops $L$ and Minkowski space $\M=P/L$, we obtain that the 
representation $\al(L)$ describes a gauge field and the induced 
representation $U(P)=\al(L)\uparrow P$ presents a particle in this 
field. We shall illustrate this approach starting from the usual 
description of a gauge field by vector-potential. 

Making use of a vector-potential $A_{\mu}(x)$, introduce a 
representation of the groupoid of pinned paths by ordered exponentials 
$$
\hat{\al}(\hat{p})={\cal P}
\exp \left\{ i\int_{\hat p} A_{\mu}(x)dx^{\mu} \right\}
$$
(integration here is performed along any of the curves from the class 
$\hat p$). Fixing an (arbitrarily) point $\Ocal\in\M$ as an origin of 
$\M$, we may associate a pinned paths $p_{\Ocal}$ (starting in 
$\Ocal$) with any free path $p\in P$. In these notations, the 
representation of loops may be expressed as 
$\al(l)=\hat{\al}(l_{\Ocal})$ and the expansion of this representation 
onto the whole group $P$ (with the properties specified above) as 
$\al(p)=\hat{\al}(p_{\Ocal})$. 

This determines a local form of the representation 
$U(P)=\al(L)\uparrow P$ which is given by the following 
elegant formula: 
$$
(U(p)\psi)(x)=\hat{\al}(p_{x'}^x)\psi(x'). 
$$
Another very convenient form of the same (or rather equivalent) 
representation may be expressed in terms of only free paths: 
$$
U(p)={\cal P}
\exp \left\{ -\int_p d\xi^{\mu}\nabla_{\mu} \right\} 
\quad \mbox{where} \quad
(\nabla_{\mu}\psi)(x)=\left( \frac{\partial}{\partial x^{\mu}} 
- iA_{\mu}(x) \right).
$$
The covariant derivatives are therefore generators of the 
representation $U(P)$ of Path Group (just as ordinary derivatives are 
generators of translations). By this the {\it group-theoretical 
interpretation of covariant derivatives} is given.

The representation of the group of loops, $\al(L)$, provides in this 
scheme a non-local description of a gauge field. This description is 
better than local descriptions by vector-potential and by 
field stregth. Indeed, the first of these descriptions is redundant 
since various vector-potentials may correspond to the same physical 
situation, and the second is insufficient because non-local 
topological effects (such as Aharonov-Bohm effect) cannot be described 
by field strength. The `intermediate' non-local description by 
$\al(L)$ is adequate. 

The scheme presented here leads to gauge fields without usage of gauge 
transformations and the idea of gauge invariance. {\it Gauge 
transformations} arise in this scheme as presentation of {\it 
arbitrariness in the local description} of gauge fields by 
vector-potentials. Let us change the vector-potential $A_{\mu}$ to 
$A'_{\mu}$ in such a way that the representation of $L$ does not 
change, $\al'(l)=\al(l)$, i.e. the non-local path-dependent 
description of the gauge field is the same. Then the new and old 
vector-potentials may be shown to be connected by a gauge 
transformation. 

\subsection{Non-Abelian Stokes theorem}

One more evidence of natural character of the non-local 
path-dependent presentation of gauge fields is the non-Abelian version 
of Stokes theorem \cite{MBM79,MBM83}. 

Any loop may be presented as a {\it product of small `lasso'} 
of the form $p^{-1}\, {\delta l}\, p$ with ${\delta l}$ being a very small loop 
and $p$ finite path (see Figure). 
\nopagebreak\begin{figure}[ht]
\let\picnaturalsize=N
\def\picsize{1.5in}
\ifx\nopictures Y\else{\ifx\epsfloaded Y\else\input epsf \fi
\let\epsfloaded=Y
{\hspace*{\fill}
 \parbox{1.5in}{\ifx\picnaturalsize N\epsfxsize \picsize\fi  
\epsfbox{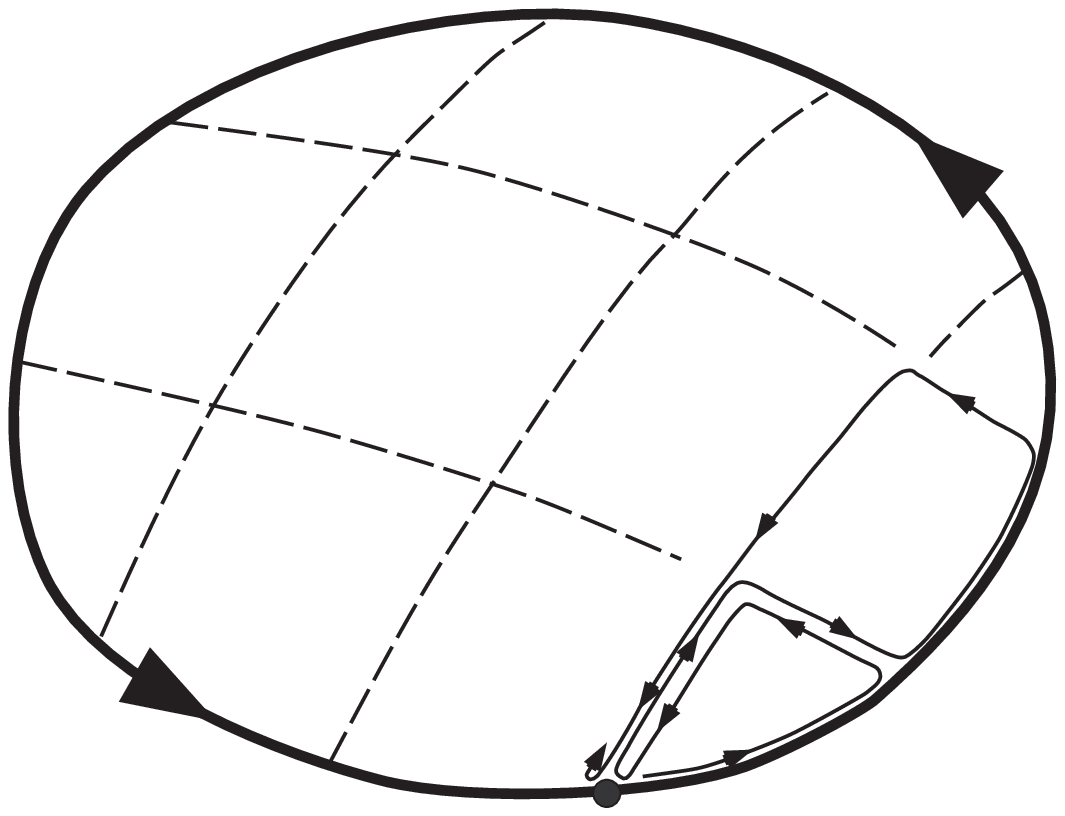}}\hfill
 \parbox{1.5in}{\ifx\picnaturalsize N\epsfxsize \picsize\fi  
\epsfbox{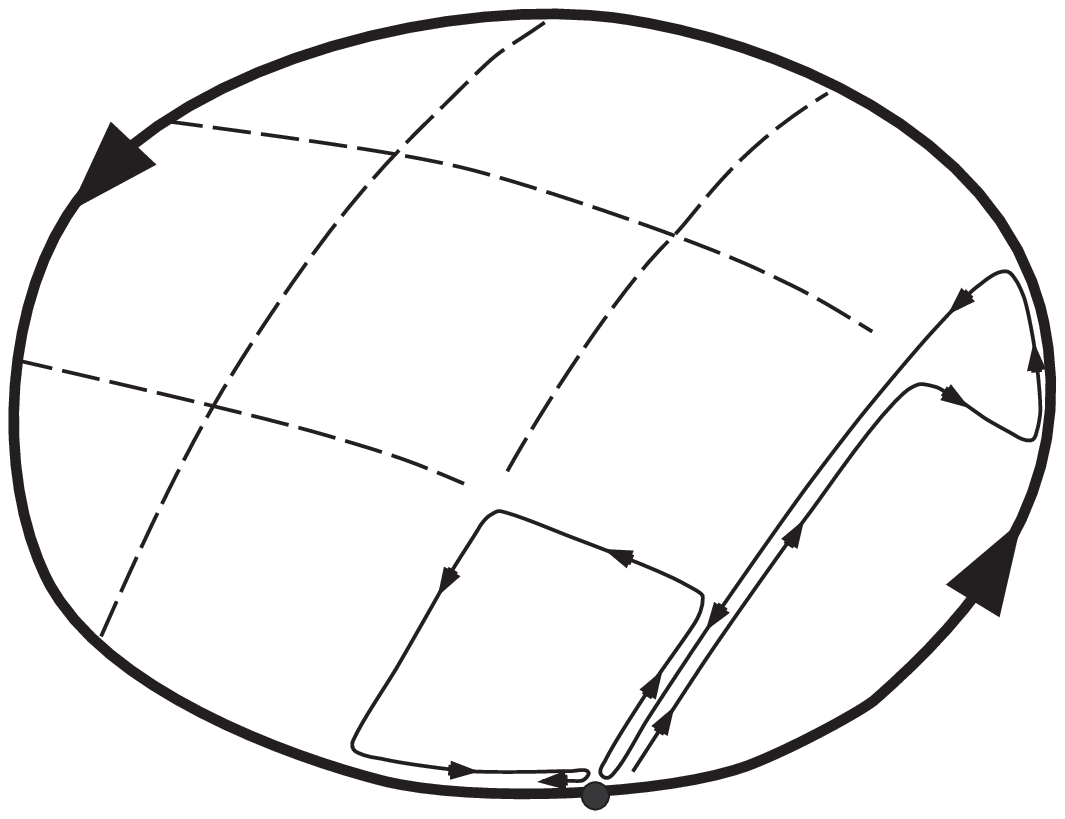}}\hfill
 \parbox{1.5in}{\ifx\picnaturalsize N\epsfxsize \picsize\fi  
\epsfbox{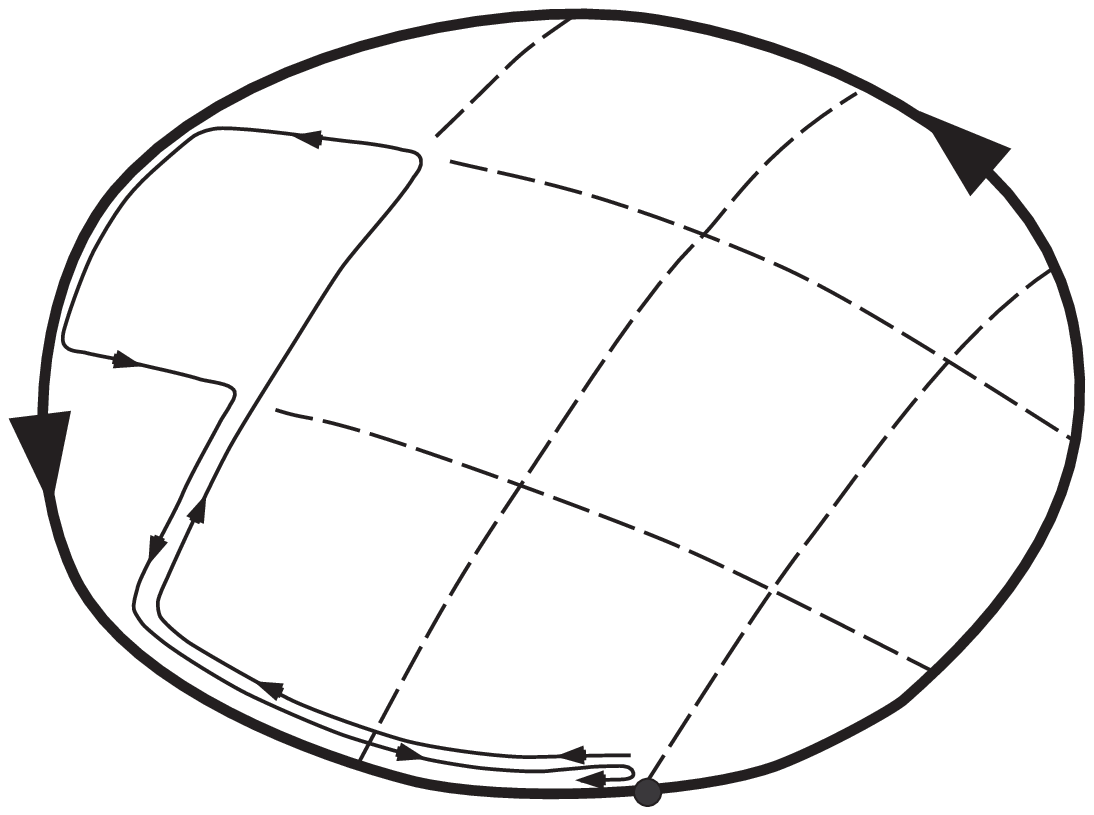}}\hspace*{\fill}}}\fi\\
\label{Fig-Stokes}
\end{figure}
The product is then of the form 
$l={\cal P} \prod_j p_j^{-1}\, {\delta l}_j \, p_j$. 
From the multiplicative properties of the representation 
$\al(L)$ we have
$$
\al(l)={\cal P} \prod_j [\al(p_j)]^{-1}\hat{\al}
\left( ({\delta l}_j)_{x_j}\right) \al(p_j) 
     = {\cal P} \prod_j 
     \exp \left( \frac{i}{2}{\cal F}_{\mu\nu}(p_j) 
     \sigma_j^{\mu\nu}\right) 
$$
where $x=p\, {\cal O}$ is a point which the path $p$ brings the origin 
$\Ocal$ to, ${\delta l}_{x}$ is a pinned loop having the shape 
${\delta l}$ and starting in $x$, 
and ${\cal F}_{\mu\nu}(p)=[\al(p)]^{-1}F_{\mu\nu}(p{\cal O})\al(p)$.
Symbolically this may be presented in the form 
$$
{\cal P} \exp \left(i\int_{\partial\Sigma}A_{\mu}(x)\, dx^{\mu}  \right) 
={\cal P} \exp 
\left(\frac{i}{2}\int_{\Sigma}F_{\mu\nu}(x)\,
dx^{\mu}\wedge dx^{\nu}  \right) 
$$
which is an ordered-exponential form of the non-Abelian Stokes 
theorem. Other forms of this theorem are in 
\cite{Stokes-other,Stokes-other2}.

\newpage

\section{Paths in gravity}

Path Group may be used for description of a gravitational field 
(curved space-time) and particles in a gravitational or gauge + 
gravitational field \cite{MBM83,Warsaw}. Minkowski space plays the 
role of a standard tangent space, connected with the curved space-time 
in a nonholonomic way. 

\subsection{Flat models of curves and fiber bundle of frames}

There is a natural (but non-holonomic) mapping 
\cite{MBM72,MyEquivPrinc} of the curves (paths) in the tangent space 
onto the curves in the curved space-time $\X$ (see Figure). 
\begin{figure}[ht]
\let\picnaturalsize=N
\def\picsize{2in}
\def\picfilename{Flat-mod.eps}
\ifx\nopictures Y\else{\ifx\epsfloaded Y\else\input epsf \fi
\let\epsfloaded=Y
\centerline{\ifx\picnaturalsize N\epsfxsize \picsize\fi \epsfbox{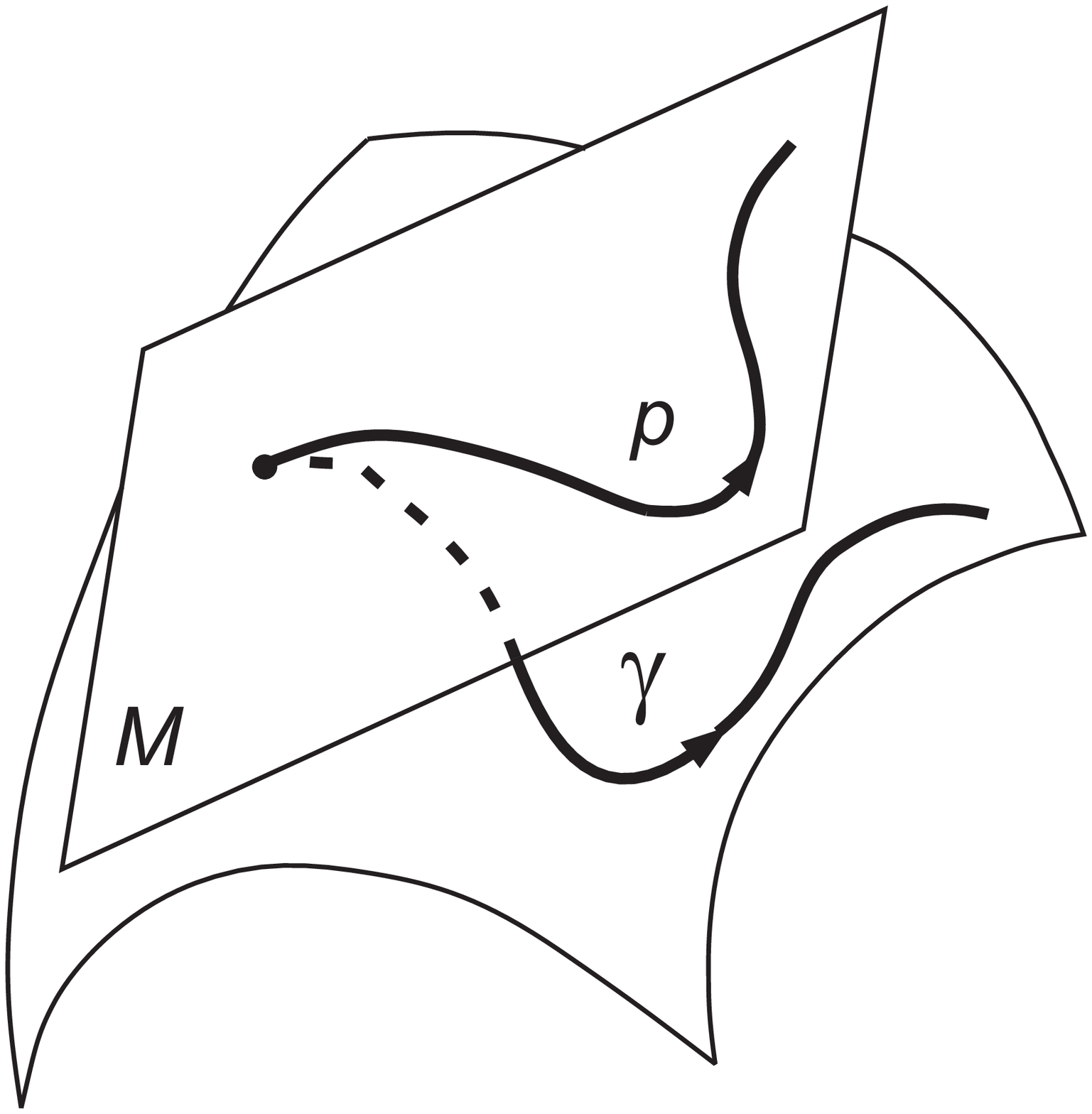}}}\fi
\label{Fig-flat-mod}
\end{figure}
The necessary relations may be presented in terms of the fiber bundle 
${\cal B}$ (over $\X$) consisting of local frames i.e. bases $b=\{ 
b_{\al}| \al=0,1,2,3\}$ of tangent spaces in the points $x\in\X$. 
Coordinates in ${\cal B}$ are $\{x^{\mu}, b_{\al}^{\mu}\}$. 

The key instrument for this end are (horizontal) {\it basis vector 
fields} in ${\cal B}$:
$$
B_{\al}=b_{\al}^{\mu}
\left( 
\partderiv{}{x^{\mu}}
-\Gamma_{\mu\nu}^{\lambda}(x)\, 
b_{\beta}^{\nu}\partderiv{}{b_{\beta}^{\lambda}} 
\right)
$$
The {\it representation of Path Group} by operators acting on 
functions in ${\cal B}$ as well as the {\it action of $P$ on 
${\cal B}$} are readily defined:
$$
U(p)={\cal P}\exp\left( \int_p d\xi^{\al}\, B_{\al}  \right), \quad 
(U(p)\psi)(b)=\psi(bp).
$$
Thus defined mapping $p: \; b\rightarrow bp$ is a parallel transport 
along the curve in $\X$ which corresponds to the path $p$ in $\M$ in 
the sense of the above mentioned mapping.

If the connection conserves the metric, these operatins may be 
restricted on the {\it subbundle $\N\in {\cal B}$ of orthonormal 
frames} yielding operators $U(p)$ acting on functions in $\N$ and the 
mapping $p:\; n\rightarrow np$ such that 
$$
(U(p)\psi)(n)=\psi(np).
$$

Thus, the action of Path Group $P$ by {\it parallel transports} is 
defined on ${\cal N}$. Lorentz group $\Lambda$ acts on ${\cal N}$ as 
the {\it structure group} of the fiber bundle, i.e. $\lambda:\; 
n\rightarrow n\lambda$ where 
$(n\lambda)_{\al}^{\mu}=n_{\beta}^{\mu}\lambda_{\al}^{\beta}$. 
Therefore the action of their semidirect product (generalized 
Poincar\'{e} group $Q=\Lambda \sd P$) is defined. 

More precisely, the elements of $Q$ are of the form $q=p\lambda$ 
(where $p\in P$ and $\lambda\in\Lambda$), multiplication of them is 
determined by the relation $\lambda [\xi ] \lambda^{-1}=[\lambda \xi 
]$, while the action of $Q$ on fiber bundle $\N$ is defined as $nq=n\, 
p\lambda = (np)\lambda$ ($p:\; n\rightarrow np$ being a parallel 
transport and $\lambda:\; n\rightarrow n\lambda$ the action of the 
structure group). The resulting action of $p\lambda$ 
(Figure, left diagram) is in accord with the multiplication law 
in the group $Q$. 
\begin{figure}[ht]
\let\picnaturalsize=N
\def\picsize{2in}
\ifx\nopictures Y\else{\ifx\epsfloaded Y\else\input epsf \fi
\let\epsfloaded=Y
{\hspace*{\fill}
 \parbox{2in}{\ifx\picnaturalsize N\epsfxsize \picsize\fi  
\epsfbox{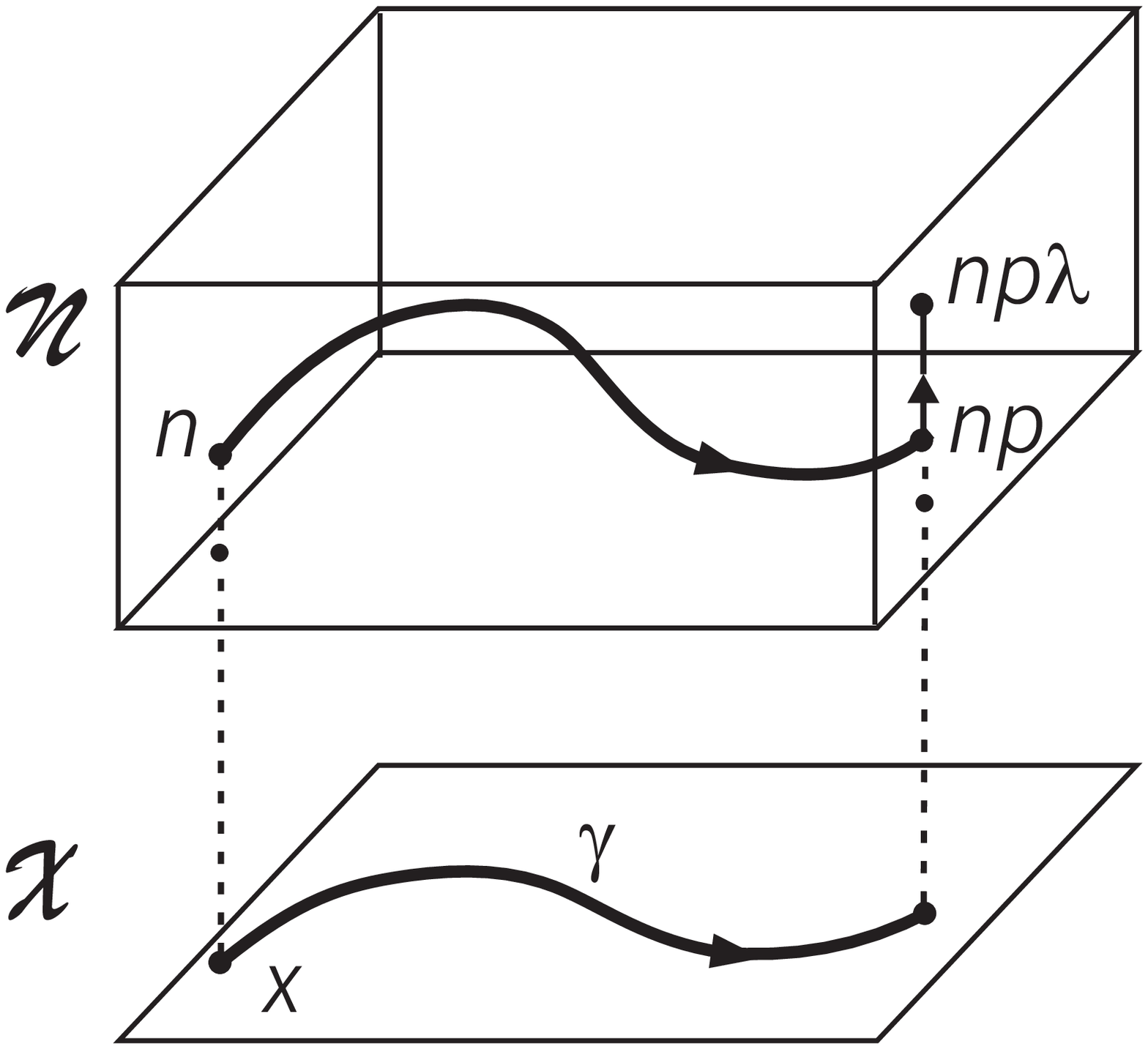}}\hfill
 \parbox{2in}{\ifx\picnaturalsize N\epsfxsize \picsize\fi  
\epsfbox{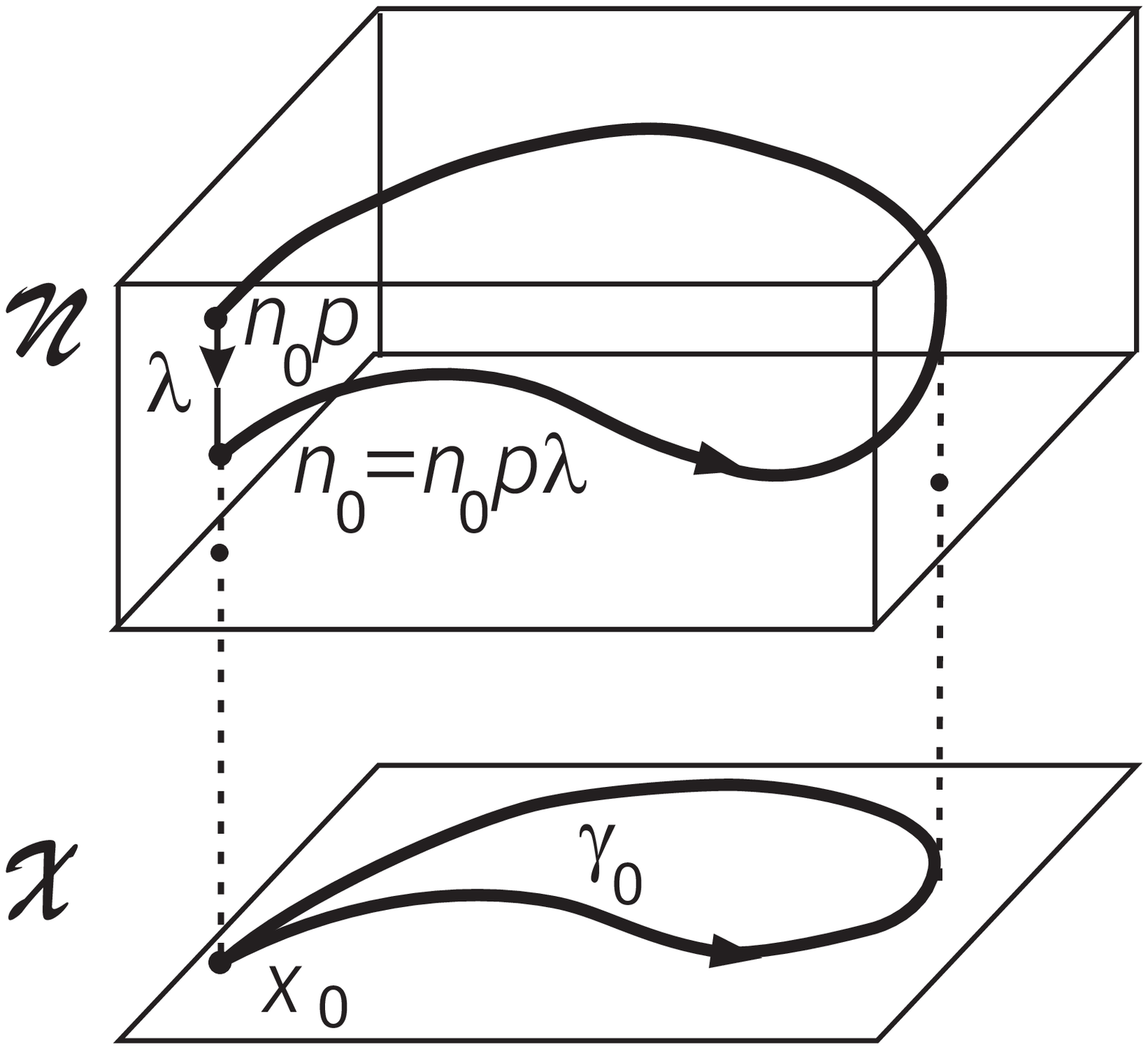}}\hspace*{\fill}}}\fi\\
\hspace*{\fill}$\mbox{Action of Gen. Poincar\'{e} Group}$\hfill \quad 
\hfill $\mbox{Element of Holonomy Subgroup}$\hspace*{\fill}\\
\label{Fig-Poincare-act}
\end{figure}

\subsection{Holonomy Subgroup}

Holonomy Subgroup $H$ of the generalized Poincar\'{e} group $Q$ is 
defined in respect to the (arbitrarily chosen and fixed) local frame 
$n_0\in\N$ as the subset leaving $n_0$ invariant (a stationary 
subgroup of $n_0$) so that $h=p\lambda\in H$ if $n_0\, p\lambda=n_0$ 
(see Figure, right diagram). 

Holonomy Subgroup $H$ in the generalized Poincar\'{e} group $Q$ plays 
the role (for gravity) analogous to the subgroup of loops $L$ in the 
Path Group $P$ (for gauge theory). This is why a representation 
$\al(H)$ represents a {\it gauge + gravitational field} while the 
induced representation $U(Q)=\al(H)\uparrow Q$ describes particles in 
this combined field. If the representation $\al$ is trivial, 
$\al(h)\equiv 1$, the induced representation describes particles in a 
pure gravitational field (corresponding to the Holonomy Subgroup 
$H$). 

Holonomy Subgroup $H\subset Q$ determines geometry. The geometry 
(including topology) may be restored if the subgroup $H$ is given. 
This reconstruction procedure may be applied \cite{LorentzCone} 
to explore geometry of hyperbolic, or Lorentzian, cones, 
starting from $H=\{ l\nu^k | l\in L, k\in Z \}$
where $\nu$ is a fixed element of the Lorentz group. 

\subsection{Quantum Equivalence Principle}

Various quantum analogues of the Einstein's Equivalence Principle 
may be defined. Validity of the quantum equivalence principle (QEP) 
depends of course on its definition. It is advantageous however to 
define QEP in such a way that it be valid. Path Group (together with 
the natural non-holonomic mapping of the curves in Minkowsky space 
onto the curves in the curved space-time) makes this possible 
\cite{MBM72,MBM74,MyEquivPrinc} (see also \cite{Pazma,Kleinert}). 

A very brief formulation of QEP may be given as follows: evolution of 
a quantum particle in a curved space-time $\X$ must be described by 
the {\it same Feynman path integral} as in the flat space-time, if the 
paths in the standard tangent space $\M$ are used in the integral 
instead of the corresponding curves in the curved space-time $\X$. 

Technically it is convenient to express QEP in terms of the 
representation $U(P)$ introduced above, putting $U(P)$ in the 
integrand of (flat) Feynman path integral instead of the 
ordinary translation. This may be done not only for purely 
gravitational but also for gravitational + gauge field (see 
\cite{MyEquivPrinc} for details). 


\section{Conclusion}

In the present paper we considered Path Group $P$ (which generalizes 
translation group) and demonstrated the following issues: 

1)~Gauge fields are described by representations $\al(L)$ of the 
subgroup $L\subset P$ of loops. Particles in such a field are 
presented by the induced representation $U(Q)=\al(L)\uparrow Q$. 
Non-Abelian Stokes theorem is naturally formulated and proved 
in terms of the representation $\al(L)$. 

2)~For application to gravity the generalized Poincar\'e group 
$Q$ is necessary which is a semidirect product of Path Group by 
Lorentz group. 

3)~Geometry of a curved space-time (including non-trivial topology) 
is presented by the Holonomy Subgroup $H\in Q$. 

4)~Gauge + gravitational field is presented by the Holonomy 
Subgroup $H$ together with its representation $\al(H)$ while 
particles in this field are described by the 
representation $\al(H)\uparrow Q$. 

5)~Quantum Equivalence Principle is naturally 
formulated in terms of Feynman path integral and 
natural non-holonomic mapping of curves in Minkowski 
space onto the curves in the curved space-time. 


\end{document}